# Traceable Coulomb Blockade Thermometry

O. Hahtela[1], E. Mykkänen[1], A. Kemppinen[1], M. Meschke[2], M. Prunnila[1], D. Gunnarsson[1,4], L. Roschier[3,4], J. Penttilä[3], J. Pekola[2]

[1] VTT Technical Research Centre of Finland, P.O. Box 1000, 02044 VTT, Espoo, Finland
[2] Low Temperature Laboratory, Department of Applied Physics, Aalto University School of Science, P.O. Box 13500, 00076 Aalto, Espoo, Finland
[3] Aivon Oy, Valimotie 13A, 00380 Helsinki, Finland
[4] D.G. and L.R. current address: BlueFors Cryogenics Oy Ltd, Arinatie 10, 00370 Helsinki, Finland

e-mail: ossi.hahtela@vtt.fi

**Abstract**

We present a measurement and analysis scheme for determining traceable thermodynamic temperature at cryogenic temperatures using Coulomb blockade thermometry. The uncertainty of the electrical measurement is improved by utilizing two sampling digital voltmeters instead of the traditional lock-in technique. The remaining uncertainty is dominated by that of the numerical analysis of the measurement data. Two analysis methods are demonstrated: numerical fitting of the full conductance curve and measuring the height of the conductance dip. The complete uncertainty analysis shows that using either analysis method the relative combined standard uncertainty ($k = 1$) in determining the thermodynamic temperature in the temperature range from 20 mK to 200 mK is below 0.5 %. In this temperature range, both analysis methods produced temperature estimates that deviated from 0.39 % to 0.67 % from the reference temperatures provided by a superconducting reference point device calibrated against the Provisional Low Temperature Scale of 2000.

**1. Introduction**

The SI-unit of thermodynamic temperature, the kelvin, will be redefined by fixing an exact numerical value of the Boltzmann constant $k_B$ presumably in 2018 [1,2]. A Coulomb blockade thermometer (CBT) is a $k_B$-based primary thermometer that can be used for a direct and accurate measurement of the thermodynamic temperature at cryogenic temperatures [3]. Such thermometers are increasingly needed in industry, research and science for practical temperature measurements, calibrations and also for defining future temperature scales, direct realization of the kelvin and dissemination of thermodynamic temperature [4].

At present, the usable temperature range for CBTs spans over more than four decades in temperature. In the millikelvin range, recent advances in cooling and on-chip thermalization of nanoscale devices enable good thermal contact between the electrons in the nanostructure and the surrounding cold bath resulting in CBT electron temperatures even below 4 mK [5]. At higher temperatures, it has been experimentally demonstrated that sophisticated electron beam lithography techniques allow for fabricating such uniform arrays of sufficiently small tunnel junctions that are suitable for accurate Coulomb blockade thermometry up to 60 K [6].

The CBTs have been compared earlier [7,8] to the Provisional Low Temperature Scale of 2000 (PLTS-2000) [9]. In [7] the agreement between the CBT and PLTS-2000 was better than 1 % in the temperature range from 0.05 K to 0.4 K. Similar results were reported in [8], where the agreement was about 1 % in the temperature range from 0.25 K to 0.65 K. Although several experiments have shown that CBT measurements agree reasonably well with other realizations of thermodynamic temperature over a wide temperature range [5,7,8,10], no thorough uncertainty analysis has been reported so far leaving the question of the traceability of the CBT measurements open.

This work presents a new CBT measurement scheme utilizing two sampling digital voltmeters (DVM). A complete uncertainty analysis demonstrates for the first time the true traceability of the CBT measurements to the SI-unit kelvin. In addition, the temperature measurement results were verified by comparing a CBT and a superconducting reference device (SRD) [11], which is directly calibrated against the PLTS-2000 in the temperature range from 20 mK to 200 mK.

## 2. Operating principles of CBT

The operation of the CBTs is based on single electron charging effects in arrays of tunnel junctions between normal metal electrodes. The interplay between thermal and charging effects can be seen as temperature-dependent changes in conductance over the tunnel junction array. The conductance curves can be numerically calculated according to the full tunnelling model presented in [3].

CBTs are classically operated in a weak Coulomb blockade regime, $E_C \ll k_BT$, where $E_C$ is single electron charging energy and $T$ is thermodynamic temperature. The charging energy of a system with $N$ junctions in series is determined as $E_C = [(N-1)/N]e^2 / C$, where $e$ is the electron charge and $C$ is the total capacitance of an island between the junctions. A characteristic nearly bell-shaped conductance curve is obtained by measuring the differential conductance, $G(V_{dc}) = dI / dV$, as a function of bias voltage ($V_{dc}$) and normalizing it against the asymptotic conductance, $G_T$, at high bias voltages (figure 1a). The normalized differential conductance dip height is determined as $\Delta G/G_T = (G_T - G_0)/G_T$, where $G_0$ is the conductance at zero bias.

In the strict limit of $E_C \ll k_BT$, the thermodynamic temperature and the full width at the half minimum of the differential conductance dip $V_{½}$ for a uniform array with mutually identical junctions are related as [3]

$$V_{½} \cong \frac{5.439 N k_B T}{e} . \qquad (1)$$

At lower temperatures, when $E_C \sim k_BT$, a linear correction term need to be included [12,13] and Equation 1 becomes

$$V_{½} \cong \left(\mathbf{1 + 0.3921}\frac{\Delta G}{G_T}\right)\frac{5.439 N k_B T}{e} . \qquad (2)$$

The linear correction depends only on the measured $\Delta G/G_T$ and therefore it does not affect to the primary nature of the CBT.

Alternatively, CBTs can be operated in a secondary temperature measurement mode in which the differential conductance dip height at zero bias voltage relates to the thermodynamic temperature in the strict limit of $E_C \ll k_BT$ as

$$\frac{\Delta G}{G_T} = \frac{1}{6} u_N , \qquad (3)$$

where $u_N = E_C/k_BT$. In this secondary mode, the temperature measurement depends on the device parameter $E_C$ which needs to be determined first, e.g., by measuring the full conductance curve at least once. Temperature measurement is much faster in the secondary mode once the $E_C$ is known. Also the self-heating effect is reduced because the conductance dip height is measured only at zero bias voltage. When reaching the intermediate Coulomb blockade regime at lower temperatures, $E_C \sim k_BT$, Equation 3 needs to be extended by higher order corrections [12] as

$$\frac{\Delta G}{G_T} \cong \frac{1}{6} u_N - \frac{1}{60} u_N^2 + \frac{1}{630} u_N^3 . \qquad (4)$$

It has been experimentally demonstrated that CBTs can be reliably operated in the intermediate temperature range if the third order corrections are employed [13].

## 3. Experimental details

### 3.1 CBT sample fabrication and preparation

The CBT sensor (figure 1b) consists of aluminium islands which are connected with tunnel junctions (figure 1c) formed by a thin $Al_2O_3$ tunnel barrier layer between two islands. The CBT samples were fabricated on an oxidized silicon substrate using an *ex situ* tunnel junction process [14]. In this method, the tunnel junctions are defined by optical lithography and plasma etching through a dielectric $SiO_x$ layer allowing for highly homogenous junction resistances in an array of tunnel junctions. The nominal diameter of the circular tunnel junctions is 0.8 μm. Additional gold thermalization blocks (figures 1d and 1e) with a thickness of ca. 5 μm were electroplated on top of the aluminium islands in order to increase the island volume and thus improve the thermal coupling between the electrons and phonons in the islands at low temperatures as described in [5]. The CBT sample used in this work consists of 20 parallel rows of 99 junctions in series. The charging energy of this design corresponds to $E_C/k_B \sim 9.9$ mK and the tunnel resistance of a single junction is $R_T \sim 23.5$ kΩ.

Excessive noise heating from the environment or through the measurement wiring can easily saturate the tunnel junction electron temperature and thus distort the CBT measurements especially at the lowest temperatures. Therefore, CBT chips contain on-chip RC-filtering structures [15] and the chips were housed in a double RF-shielded sample holder consisting of a gold-plated copper chamber with indium wire sealing. In order to further improve the filtering against high-frequency interference, thermocoax cables [16] were used as four-probe measurement lines from the sample holder to the room temperature flange of the cryostat. The thermocoax wires were also thermally anchored to the mixing chamber plate of the cryostat. Two small neodymium permanent magnets are located close to the CBT chip providing a magnetic field of a few tens of millitesla perpendicular to the chip surface to suppress the superconductivity of the aluminium islands below 1.2 K. The CBT sample holder and the SRD unit were mounted on a gold-plated copper platform attached with a weak thermal link to the mixing chamber plate of the cryostat (Figure 1f). The temperature of the platform was controlled with a resistive heater and stabilized to the midpoint of the superconductive transitions of the SRD with a PID-controller. The measurements were carried out in a commercial cryogen-free dilution refrigerator with a base temperature of ca. 15 mK and cooling power of 250 μW at 100 mK [17].

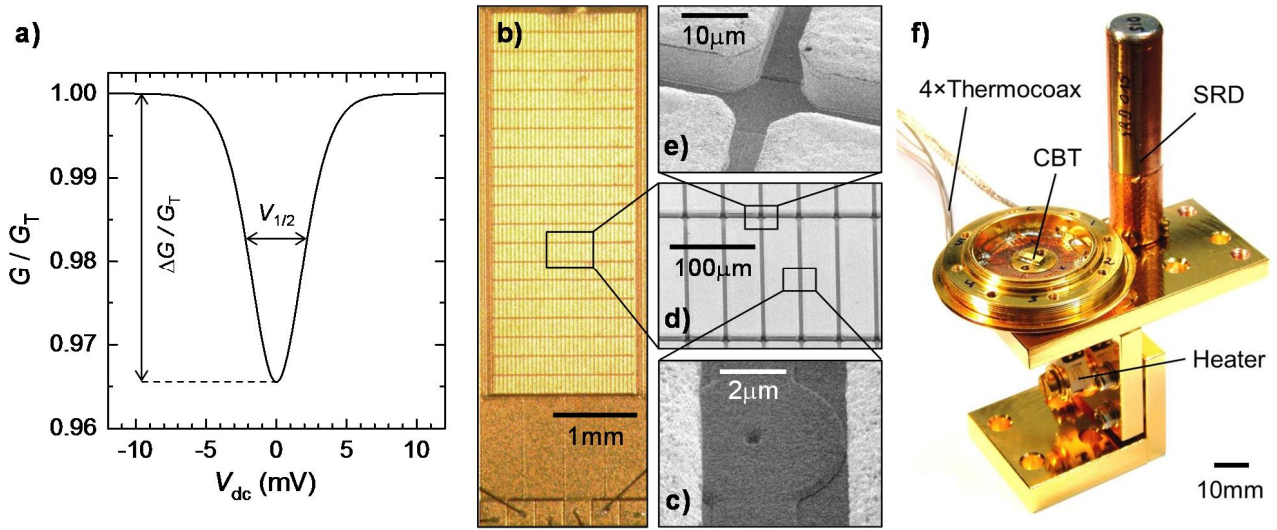

**Figure 1**. a) Normalized differential conductance as a function of dc bias voltage. b) A photograph of a CBT chip. c) A SEM image of a circular tunnel junction between two thermalization blocks. d) A SEM image of adjacent Au thermalization blocks. e) A close up of the corners of the electroplated thermalization blocks. f) CBT sample holder (shown without the indium wire sealed cover) and SRD unit are mounted on a gold-plated copper platform. The temperature of the platform is controlled with a resistive heater.

## 3.2 Measurement scheme and data acquisition

The measurement setup is shown in Figure 2a. The source voltage $V_{source} = V_{dc} + V_{ac}$ is generated as in the CBT experiments of [8]. Separate floating voltage sources and input resistors are used for producing dc bias voltage $V_{dc}$ and sinusoidal ac excitation voltage $V_{ac}$. The resistors $R_{dc}$ and $R_{ac}$ form together with the input resistor $R_{in}$ separate voltage divisions for dc and ac. Typically the bias voltage $V_{dc}$ range is set from $-3V_{½} \ldots +3V_{½}$ to $-5V_{½} \ldots +5V_{½}$ and $V_{ac}$ is only a small fraction (2 % … 5 %) of the $V_{½}$. The $V_{ac}$ is needed for the excitation to determine the differential conductance of the CBT, and its magnitude is a trade-off between sufficiently high ac output signals ($dV$ and $dI$) and small enough excitation level to reduce distortion of the conductance curve. Floating voltage supplies are used to prevent ground loops.

The body of the cryostat serves as the measurement ground. The current is amplified with a transimpedance amplifier and the voltage with a differential voltage amplifier that has isolated inputs. Both amplifiers are battery powered and external regulators stabilize the supply voltages to prevent drifts of the amplifier temperatures. The measurement lines are triaxial: A coaxial cable is used as a signal line. In addition, the coaxial cable has a copper braid shield that connects the guard shield of the multimeter to the measurement ground. In the current measurements the multimeter guard is isolated from the LO input inside the multimeter. This configuration suppresses the effect of ground loops coupled by the stray capacitances of the multimeter. In the voltage measurements, the guard is connected to the LO input to carry the measurement ground to the multimeter.

In a traditional CBT experiment, the differential conductance curve $G(V_{dc}) = dI / dV$ is measured by using two lock-in amplifiers to detect both the ac current $dI$ and voltage $dV$ signals arising from the $V_{ac}$ excitation. Both signals are amplified before the lock-in devices. The dc bias voltage can be deduced from $V_{in,dc}$ and the dc voltage division. The uncertainty of such two-probe measurement can be as low as some hundreds of ppm [8,18], but is prone to the changes in the setup. While the lock-in technique may be convenient for typical CBT measurements, we developed a new measurement scheme for metrological purposes that could also be used to validate simpler setups.

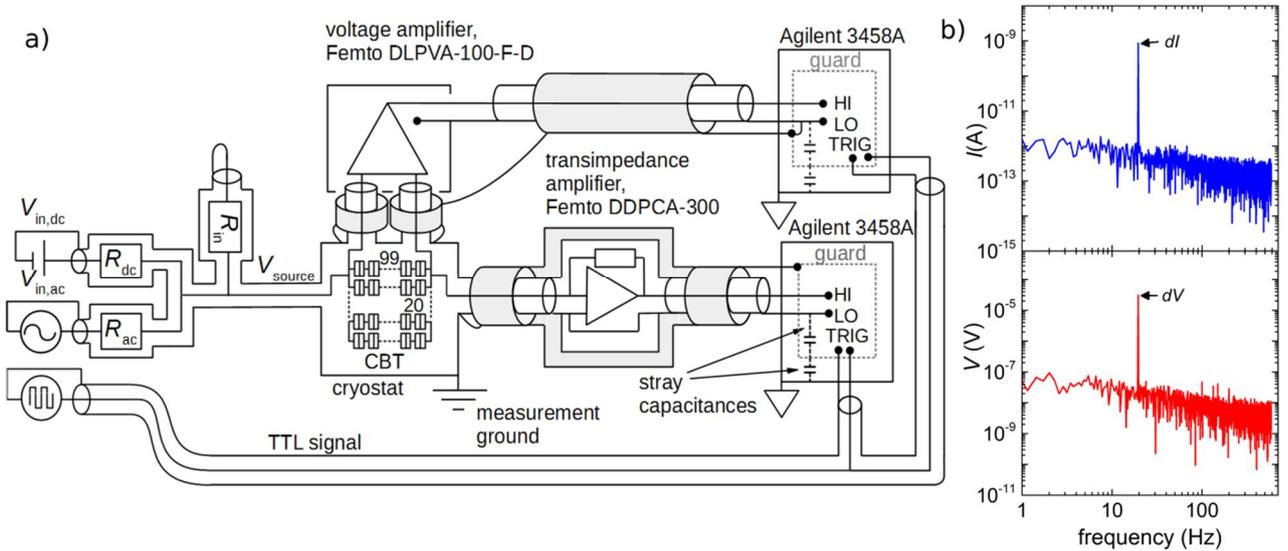

**Figure 2.** a) Measurement setup for reading the CBT is based on using two sampling digital voltmeters. See the text for the details. b) Example Fourier transforms of the digitized data. The excitation signals are single peaks, and all other disturbances are below or comparable to the noise floors of the amplifiers.

In this paper, two Agilent 3458A multimeters are used to measure the output signals of the voltage and current amplifiers. The analog-to-digital conversion (ADC) of the multimeters is triggered with a floating TTL signal that shares the same clock with the source that generates the $V_{ac}$. This allows to digitize exactly an integer number of periods of the sinusoidal excitation signal. Then the $V_{ac}$ does not affect the value of the $V_{dc}$, which is obtained from the average of the measured $V$ signal. After employing a standard Fast Fourier Transform (FFT) algorithm, the excitation shows up as single peaks in the frequency spectra of the measured $I$ and $V$ signals (figure 2b). In this work the excitation signal frequency is $f_{V,ac}$ ~ 19.5 Hz.

This technique has several benefits compared to the lock-in technique: (i) When both the voltage amplifier and the multimeter are calibrated, the dc voltage over the CBT sensor is traceable. The measurement is also four-probe. (ii) The input of the Agilent 3458A multimeter is floating, unlike those of typical lock-in amplifiers, and the guarding options effectively suppress capacitive ground loops. (iii) Our measurement software plotted the FFT spectra from 1 Hz to 1 kHz, which allowed us to always check that there was no unwanted disturbance at other frequencies. In particular, if mistakes were made with the guarding connections, disturbance peaks at the multiples of 50 Hz appeared in both spectra, and the connections could be corrected. Both the FFT and the lock-in amplifier measurement techniques are mathematically insensitive to noise at other frequencies than the excitation, but the other disturbances would heat the CBT sensor and cause a systematic error. The possible heating due to the $V_{source}$ is taken into account by the CBT analysis algorithms as discussed later.

**3.3 Conductance curve analysis and fitting**

A full tunnelling model to numerically calculate the conductance curve $G(V_{dc})$ of the CBT at any ratio of $E_C$ to $k_BT$ is presented in [3]. This model can be expanded to take into account the Joule heating effects due to the applied bias voltage [7].

We used a freely available Python-based pyCBT software [19] to analyse the measured differential conductance curves. The algorithm uses the tunnel junction resistance $R_T$, capacitance of an island $C$ and phonon temperature $T_p$ as fit parameters and it takes into account the possible heating due to the measurement itself and environment according to [7]. The validity of the temperature readings obtained from the numerical fitting were checked by inspecting that the measured and fitted conductance curves are as consistent as possible with each other. This kind of inspection is important as it was found out during our measurements and data analysis that excessive noise and ripple in the measurement signal may slightly distort the fitting procedure. This was evidenced e.g. by a slightly too low conductance dip at zero bias if compared to the measured curve which resulted in a small (typically 1-3 %) error in the $T_p$ readings.

As an example of a successful temperature analysis, figure 3 illustrates that in a measurement performed at a SRD-provided reference temperature of $T_{2000}$ = 93.08 mK, the measured (blue dots) and fitted (black line) conductance curves coincide best when the fit parameter $T_p$ is fixed to 93.71 mK. Figure 3 also shows the ratio between the measured ($G_{meas}$) and fitted ($G_{fit}$) conductances at 93.71 mK indicating that the residuals of the fitting are comparable to the noise level of the measured conductance values.

In order to demonstrate the sensitivity of the conductance curve to the temperature variations, figure 3 also shows the simulated conductance curves for the fixed temperatures $0.99 \times 93.71$ mK = 92.77 mK (blue line) and $1.01 \times 93.71$ mK = 94.65 mK (red line). It should be noted that the changes in the fitted conductance curve due to this small (one percent) temperature variations are much easier to discover from the change in the conductance dip height at zero bias, $G_0$, than from the change in the full width at half minimum, $V_{½}$.

We also used the secondary temperature measurement mode, i.e. the experimentally determined $\Delta G/G_T$ and the fitted $E_C$ values to calculate the thermodynamic temperature from the CBT measurements (Equation 4). The third order approximation was chosen to be used here as the operating temperatures covered in this work shift from the weak Coulomb blockade regime ($E_C << k_BT$ at 200 mK) towards the intermediate Coulomb blockade regime ($E_C$ ~ $k_BT$ at 20 mK). The temperature values obtained from the secondary mode measurements were also used for crosschecking the validity of the numerical fitting. The secondary mode measurement and the numerical fitting produced temperature values that agreed better than 0.05 % at all five reference temperature points from 20 mK to 200 mK.

The influence of the randomly fluctuating background charges on the measured conductance of the CBT array is difficult to resolve experimentally. Fortunately the background charges can be neglected in the temperature range covered in this work [13]. No sign of significant noise heating from the environment or due to poor thermalization that would saturate the electron temperature were observed during the CBT measurements down to 20 mK.

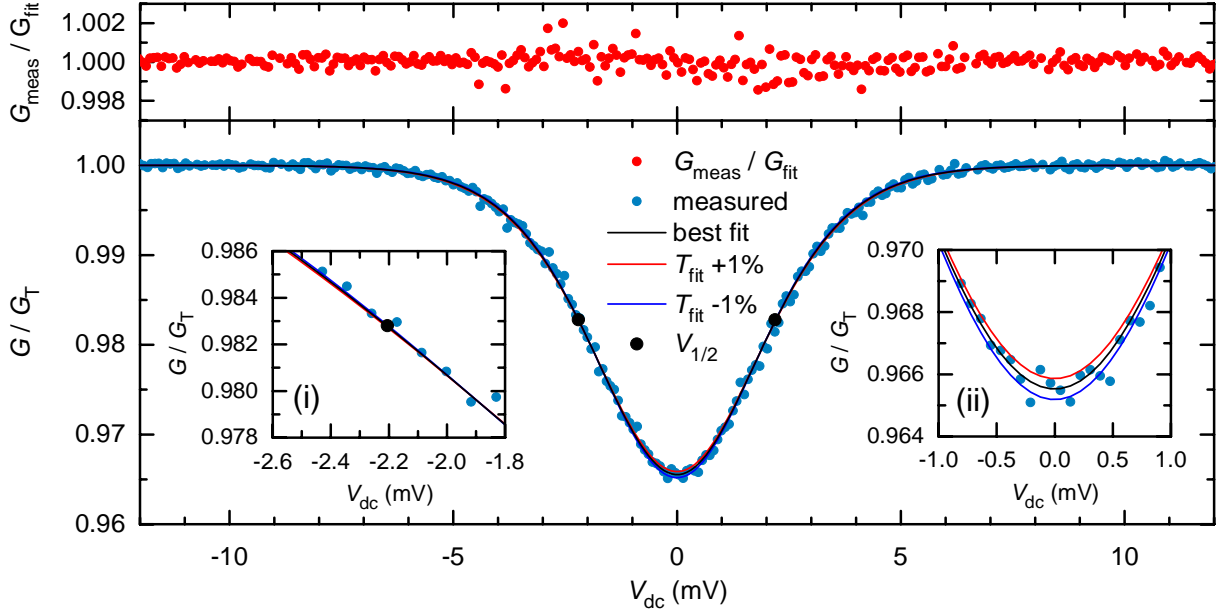

**Figure 3.** Measured and fitted normalized conductance curves at $T_{2000} = 93.08$ mK, where the best fit to the measured data gives $T_{CBT} = 93.71$ mK. Inset (i) shows the conductance curve in the vicinity of the conductance at the negative half minimum point. Inset (ii) shows a close up of the conductance dip close to the zero bias.

### 3.4 Results and verification of the temperature measurement

A superconductive reference point device SRD-1000 [11] that was earlier calibrated against PLTS-2000 at PTB, Germany was used to provide five temperature reference points from 21 mK to 209 mK (see Table 1) for checking the accuracy of the CBT measurements and consistency to $T_{2000}$ [20].

The uncertainties of the five reference temperatures provided by the SRD as well as the stabilities of the reference temperatures during the comparison measurements are given in Table 1. Also, the thermodynamic temperatures, $T_{CBT}$, obtained from the numerical fitting at five reference points and the corresponding uncertainties and relative deviations from the reference temperatures are listed in Table 1. The measured conductance curves at all five reference temperatures are shown in Figure 4. As shown in the inset of the Figure 4, it was noticed that the thermodynamic temperatures obtained from the SRD and CBT measurements agree within the expanded measurement uncertainty ($k = 2$) but CBT shows systematically (0.39 %...0.67 %) higher temperatures than $T_{2000}$. The reasons behind the systematic difference were not resolved in this work but require more research about the possible inherent errors related to the CBT sample and measurement and analysis techniques.

**Table 1.** Thermodynamic temperatures were measured with the CBT sample at five $T_{2000}$ reference points determined by a superconductive reference point device SRD that was calibrated against PLTS-2000. The uncertainties $u(T_x)$ are combined standard uncertainties with coverage factor $k = 1$.

| $T_{2000}$ mK | $u(T_{2000})$ mK | $T_{2000}$ stability mK | $T_{CBT}$ mK | $u(T_{CBT})$ mK | $u(T_{CBT})$ % | $(T_{CBT}-T_{2000})/T_{2000}$ % |
|---|---|---|---|---|---|---|
| 20.73 | 0.06 | 0.004 | 20.83 | 0.09 | 0.45 | 0.50 |
| 65.58 | 0.24 | 0.007 | 65.91 | 0.32 | 0.48 | 0.50 |
| 93.08 | 0.16 | 0.016 | 93.71 | 0.45 | 0.48 | 0.67 |
| 155.47 | 0.12 | 0.011 | 156.08 | 0.76 | 0.49 | 0.39 |
| 207.78 | 0.20 | 0.014 | 208.64 | 1.00 | 0.48 | 0.41 |

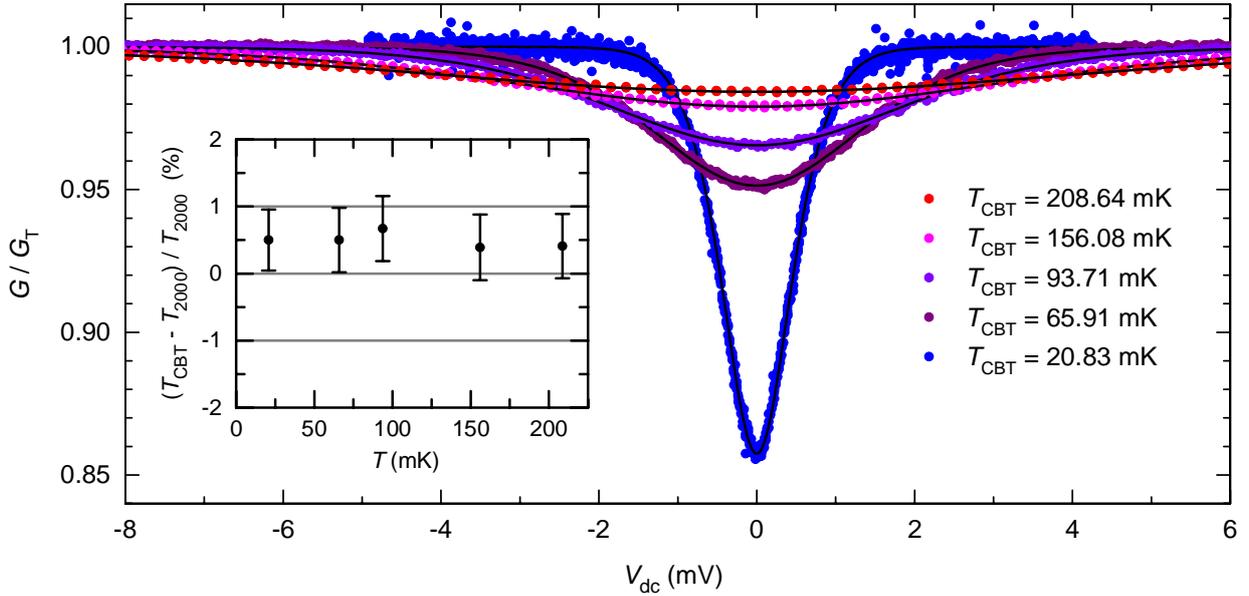

**Figure 4.** Measured and fitted normalized conductance curves from 21 mK to 209 mK. Inset shows the relative difference between the temperatures given by SRD and CBT. $T_{CBT}$ refers to the temperatures obtained from the numerical fitting of the measured conductance curves. The error bars denote combined standard uncertainties ($k = 1$).

## 4. Measurement uncertainty analysis

### 4.1 Primary mode analysis

The uncertainty analysis of the thermodynamic temperature measurement using CBTs was carried out according to the *Evaluation of measurement data - Guide to the expression of uncertainty in measurements* (GUM) [21] and discussions presented in [18]. Table 2 presents an example of an uncertainty budget for the CBT measurement performed at $T_{CBT} = 93.71$ mK. The major contributions to the uncertainty of the CBT measurements using the numerical fitting of the measured conductance curves were identified as follows:

$u(V_{dc,cal})$  The accuracy and correctness of the bias voltage $V_{dc}$ measurement determines how well the width of the conductance dip can be determined. Therefore, the DVM and the gain of the voltage preamplifier need to be calibrated traceably. The dominating contribution to the uncertainty of the dc voltage measurement originates from the uncertainty in the voltage preamplifier gain calibration which was 55 ppm ($k = 1$) for the nominal gain of 40 dB and voltage range relevant in this work.

| $u(V_{dc,stab})$ | The standard uncertainty due to the long term drift of the gain was measured to be 20 ppm ($k = 1$) by calibrating the voltage preamplifier gain before and after the CBT measurements. |
| --- | --- |
| $u(V_{dc,t})$ | The gain of the voltage preamplifier depends slightly on the ambient temperature. The temperature dependence of the gain was determined to be -30 ppm/°C at room temperature. The ambient temperature in the temperature controlled laboratory room stays within ±1 °C from the calibration temperature during the CBT measurements. |
| $u(R_T)$ | The full tunnelling model requires that all the tunnel junctions in a CBT array are identical. In practice, the junction parameters, such as junction areas and tunnel barrier thickness, may experience some variance due to the small variations in the fabrication process. Deviation in the junction resistances $R_T$ makes the conductance dip appear narrower than that given by an ideal homogeneous CBT array thus giving lower temperature readings [22]. The *ex situ* fabrication method allows for fabricating highly homogenous tunnel junctions. According to the studies reported in [14] we assume that the local deviation in the tunnel junction resistance values of the CBT samples used in this work is less than 2 %. The error in temperature due to the fabrication inhomogeneity can be estimated from $$\frac{\delta T}{T} \approx -k\sigma^2 , \qquad (5)$$ where the numerical factor $k \sim 0.73+(N-1)/N$ and $\sigma = \delta R/R_0$ is the rms deviation of the junction resistances relative to the mean resistance value $R_0 = R_\Sigma/N$, where $R_\Sigma$ is the sum of all junction resistances in series [22]. Thus, we estimate that the standard uncertainty due to the junction resistance variations in our CBT measurements is less than 0.07 % ($k = 1$). |
| $u(T_{fit})$ | The precondition for assessing if the numerical fitting of the conductance curve is successful is that the fitted and the measured conductance curves are uniform and as consistent as possible and especially that the conductance dip heights are the same. This was estimated by calculating the residuals from the fitted and measured conductances. The relative deviation of the residuals compared to the conductance dip height determines the uncertainty due to the numerical fitting. In our measurements, this uncertainty component was determined to be about 0.4 % ($k = 1$) at each measured temperature point. |
| $u(T_{rep})$ | The repeatability of the CBT measurements was determined by measuring a few conductance curves at each fixed reference temperature. Because the number of the repeated measurements was rather low due to the limited time resources, the statistical uncertainty has quite significant contribution, 0.25 % ($k = 1$), to the total uncertainty. |

**4.2 Secondary mode analysis**

The uncertainty analysis for the secondary mode temperature measurements according to the Equations 3 and 4 differs only slightly from that of the numerical fitting method because the full conductance curve needs to be measured at least once for determining the charging energy $E_C$ of the CBT sample. In the secondary mode, the uncertainty components related to the determination of the conductance dip height $u(\Delta G/G_T)$ and charging energy $u(E_C)$ together replace the $u(T_{fit})$ component in the uncertainty analysis presented above.

| $u(\Delta G/G_T)$ | The depth of the normalized conductance dip $\Delta G/G_T = (G_T - G_0)/G_T$ is determined from the asymptotic conductance value $G_T$ and conductance value at zero bias voltage $G_0$. Because the measured conductance values are normalized against the $G_T$ and we are mainly interested in the shape of the conductance curve, the stability of the differential conductance measurement during the bias voltage sweep is more critical than the absolute accuracy of the measured conductance values. This uncertainty component varied between 0.14 %…0.34 % ($k = 1$) at different temperatures and it takes into account the uncertainties in determining both the $G_T$ and $G_0$ individually. |
| --- | --- |

| $u(E_C)$ | The determination of the charging energy $E_C$ of a CBT sample was noticed to introduce a large contribution, 0.25 % ($k = 1$), to the total uncertainty budget of the secondary mode measurement. The $E_C$ value received from the numerical fitting is quite sensitive to the fitting input parameters and possible overheating effects but the dominating contribution here comes from the variance in the $E_C$ values determined at different temperatures. |
|---|---|

The relative combined standard measurement uncertainty ($k = 1$) using the numerical fitting varied between 0.45 % and 0.49 % in the temperature range from 20 mK to 200 mK while the corresponding relative combined standard uncertainties for the secondary mode measurements varied from 0.38 % to 0.49 %. As expected, the relative uncertainty values are close to each other because the uncertainty analysis for both numerical fitting and secondary mode are based on the same measurement data and, due to the constant relative $0.02 \times V_\frac{1}{2}$ ac excitation, the signal to noise ratio in the measured conductance curves is about the same at each measured temperature.

In our experiments, the measurement of a full conductance curve for collecting data for the numerical fitting took typically about 4 hours. The temperature measurement using a CBT in a secondary mode is essentially much faster, because, once the $E_C$ and $G_T$ values have been determined, only the $G_0$ value needs to be measured at zero bias voltage to resolve the conductance dip height. Thus, instead of measuring the full conductance curve, it is easier to concentrate on determining the $G_0$ value with higher accuracy and lower uncertainty, e.g., by using even longer averaging times or collecting more data at zero bias. This also offers possibilities to improve the overall measurement uncertainty through the reduced $u(\Delta G/G_T)$ uncertainty component. In principle, the $E_C$ and $G_T$ values stay constant at different temperatures in normal operation. However, it is advised to regularly check the $E_C$ and $G_T$ values also at different temperatures when operating the CBT in the secondary mode so that the stability of the $E_C$ and $G_T$ values can be taken into account in the uncertainty analysis.

**Table 2** Uncertainty budgets for the temperature measurements based on the numerical fitting of the conductance curve, i.e. primary mode, and using the secondary mode performed at $T_{CBT} = 93.71$ mK.

| Uncertainty source | Primary mode | | Secondary mode | |
|---|---|---|---|---|
| | $u(x_i)$ / mK | $u_{rel}(x_i)$ / % | $u(x_i)$ / mK | $u_{rel}(x_i)$ / % |
| $u(V_{dc,cal})$ | 0.005 | 0.0055 | 0.005 | 0.0055 |
| $u(V_{dc,stab})$ | 0.002 | 0.002 | 0.002 | 0.002 |
| $u(V_{dc,t})$ | 0.002 | 0.002 | 0.002 | 0.002 |
| $u(R_T)$ | 0.066 | 0.07 | 0.066 | 0.07 |
| $u(T_{fit})$ | 0.383 | 0.408 | - | - |
| $u(\Delta G/G_T)$ | - | - | 0.318 | 0.339 |
| $u(E_C)$ | - | - | 0.234 | 0.25 |
| $u(T_{rep})$ | 0.234 | 0.25 | 0.234 | 0.25 |
| Combined standard uncertainty ($k$=1) | 0.45 | 0.48 | 0.46 | 0.49 |

## 5. Summary

We have developed and tested a traceable measurement scheme for determining thermodynamic temperature at low temperatures using CBTs. In this method, a traditional lock-in technique is replaced by two sampling digital voltmeters which helps to avoid problems related to the grounding of the electronics and allows for collecting more detailed data e.g. for analysing measurement noise that could overheat the electrons in the CBT arrays at low temperatures. Also the contribution of the bias voltage $V_{dc}$ measurement to the total measurement uncertainty is drastically reduced with the new technique.

We report the uncertainty analysis for temperature measurement based on the numerical fitting of the full conductance curve and for using the CBT in the secondary measurement mode. The combined standard measurement uncertainty ($k = 1$) in both cases was below 0.5 % in the temperature range from 20 mK to

200 mK. The accuracy of the CBT measurements was verified by comparing the CBT and SRD that was earlier calibrated against PLTS-2000. The results were in a good agreement within the expanded measurement uncertainty ($k = 2$) in all five temperature reference points from 20 mK to 200 mK. This work presents for the first time a CBT-based measurement scheme for determining the thermodynamic temperature that is traceable to the SI-unit kelvin.

**Acknowledgements**

This work has received funding from European Union on the basis of Decision No 912/2009/EC. The research has been carried out in the framework of the European Metrology Research Programme SIB01-Implementing the new kelvin [4]. The authors thank Ilkka Iisakka from VTT-MIKES for calibrating the voltage amplifier, Dr. Jari Hällström from VTT-MIKES for useful discussions in developing the new measurement scheme and Dr. Jost Engert and Dr. Alex Kirste from PTB, Germany for loaning the calibrated SRD device. A.K. and M.P. acknowledge the support by the Academy of Finland (grant number 259030 and project number 287768, respectively). E.M. acknowledges the support of the Wihuri Foundation.